\documentclass[a4paper,12pt]{article}       
\usepackage{latexsym}

\begin{document}

\title{The local and global geometrical aspects of the twin paradox in static spacetimes: 
I. Three spherically symmetric spacetimes}
     
\author{Leszek M. SOKO\L{}OWSKI and Zdzis\l{}aw A. GOLDA \\
Astronomical Observatory, Jagiellonian University,\\
Orla 171,  Krak\'ow 30-244, Poland,\\ 
and Copernicus Center for Interdisciplinary Studies,\\ 
email: lech.sokolowski@uj.edu.pl,\\
email: zdzislaw.golda@uj.edu.pl} 

\date{}
\maketitle

\begin{abstract}
We investigate local and global properties of timelike geodesics in three static spherically 
symmetric spacetimes. These properties are of its own mathematical relevance and provide a 
solution of the physical `twin paradox' problem. The latter means that we focus our studies on 
the search of the longest timelike geodesics between two given points. Due to problems with 
solving the geodesic deviation equation we restrict our investigations to radial and circular 
(if exist) geodesics. On these curves we find general Jacobi vector fields, determine by means 
of them sequences of conjugate points and with the aid of the comoving coordinate system and the 
spherical symmetry we determine the cut points. These notions identify segments of radial and 
circular gepdesics which are locally or globally of maximal length. In de Sitter spacetime all 
geodesics are globally maximal. In CAdS and Bertotti--Robinson spacetimes the radial geodesics 
which infinitely many times oscillate between antipodal points in the space contain infinite 
number of equally separated conjugate points and there are no other cut points. Yet in these 
two spacetimes each outgoing or ingoing radial geodesic which does not cross the centre is 
globally of maximal length. Circular geodesics exist only in CAdS spacetime and contain an 
infinite sequence of equally separated conjugate points. The geodesic curves which intersect 
the circular ones at these points may either belong to the two-surface $\theta=\pi/2$ or lie 
outside it.\\

Keywords: twin paradox, static spherically symmetric spacetimes, Jacobi fields, conjugate and 
cut points\\

PACS: 04.20Jb
\end{abstract}
 
\section{Introduction}
We provide detailed calculations concerning the 'twin paradox' problem in three particular static 
spherically symmetric (SSS) spacetimes. We consider three twins following different worldlines 
 joining common endpoints and establish which twin gets the oldest one at the reunion. As is 
 well known, the problem is of purely geometrical nature and in this setting is equivalent to the 
 search in differential Lorentzian geometry of the longest timelike curve joining two given points 
 in the spacetime. The problem actually consists of two separate problems: a local and a global 
 one. In the local problem one considers a bundle of nearby (infinitesimally close) timelike 
 curves and seeks for the longest one in the bundle. Again it is well known that there is a well 
 defined procedure for solving the local problem in terms of the curvature tensor, which 
 physically determines the behaviour of geodesic worldlines of nearby free test particles both in 
 four and in a larger number of spacetime dimensions \cite{PS}. A locally maximal timelike curve is 
 always a geodesic and is determined by solving the geodesic deviation equation. If the endpoint 
 (the reunion point of the twins' worldlines) does not lie in a convex normal neighbourhood of 
 the initial point, the two points are connected by two or more geodesics of the bundle. This fact 
 is signalled by the existence of points conjugate to the initial one lying on one of these 
 geodesics. In other terms a segment of a geodesic $\gamma$ joining points $P_0$ and $P_1$ is 
 locally of maximal length between these points if there are no conjugate points to $P_0$ on 
 $\gamma$ within the segment. All necessary theorems concerning Jacobi vector fields (the deviation 
 vectors) and conjugate points are briefly summarized in \cite{S}, where locally maximal worldlines 
 in Schwarzschild metric were studied.\\
 Yet the global problem is quite different: here one seeks for the longest curve among all possible 
 timelike ones joining the given points $P_0$ and $P_1$. This means that one compares the lengths 
 of curves which besides the endpoints are distant from each other. It is clear that the nonlocal 
 nature of the problem precludes the existence of any analytic tool to establish if the given 
 curve is globally maximal: there is no differential equation (playing the role of the deviation 
 equation) whose solutions might indicate the longest curve. The globally maximal curve is again 
 a segment of a timelike geodesic and the notion of the conjugate point is replaced by the cut 
 point indicating the end of this segment. All what is known in global Lorentzian geometry in this 
 respect are 'existence theorems' which provide no effective algorithm for searching for 
 maximal geodesics. On the contrary, in general one should consider the whole set of timelike 
 curves with the given common endpoints and compare their lengths case by case.\\
 High symmetry, such as the spherical one, may help in this search, however, as we shall see 
 below, the spacetimes with the same symmetry considerably differ from each other. Spherical 
 symmetry is singled out since in these spacetimes it is quite easy to find a transformation from 
 the coordinates in which the spacetime metric is originally given to the comoving coordinates. 
 In the latter coordinates it is straightforward to find out globally maximal segments for a 
 class of timelike geodesics.\\
 
 The purpose of the work is twofold: firstly, to find out complete sets of solutions for Jacobi 
 vector fields for two classes of timelike geodesics, determine conjugate points on them (being 
 zeros of the Jacobi fields) and in this way show the locally maximal segments of these curves, 
 then establish, where it is possible, whether these segments (or their pieces) are globally 
 maximal. Secondly, we interpret physically these geometrical properties of geodesics in the 
 framework of the twin paradox: which twin's worldline makes him the oldest one. The 
 mathematical apparatus applied to deal with the global maximality problems is described in our 
 previous paper \cite{SG1} and we refer the reader to it.\\
 We emphasize that the search for both locally and globally maximal geodesics is doubly limited. 
 Firstly, an exact analytic expression for the geodesic is necessary and since it is a solution 
 to the nonlinear system of equations, it is available only in a narrow class of spacetimes 
 having sufficiently high symmetries. Secondly, the explicit form of the geodesic is used in the 
 geodesic deviation equation, which is nonlinear in the tangent vector to the geodesic making 
 the equation quite complicated. All timelike geodesic solutions in Schwarzschild spacetime are 
 known and are given in terms of Weierstrass elliptic functions (see \cite{HL} and references 
 therein). One should not expect that the equation might be effectively solved when these 
 functions appear in it.\\
 For this reason we consider only static spherically symmetric spacetimes. For their metrics one 
 has two classes of physically distinguished and analytically simple timelike geodesics: radial 
 and circular (if exist) ones. Staticity not only considerably simplifies all calculations, 
 moreover it allows for a physically meaningful notion of rest\footnote{An unambiguous notion 
 of rest may also be defined in some time--dependent spacetimes, e.~g.~in Robertson--Walker 
 world.}. In an SSS spacetime we introduce three twins: twin A remains at rest on a 
 non--geodesic worldline, twin B revolves on a circular orbit (geodesic or not) around the centre 
 of spherical symmetry and twin C moves upwards and 
 downwards following a radial geodesic. The twins' worldlines emanate from a common initial 
 point and we study under what conditions they will intersect in the future. We make detailed 
 calculations in three SSS spacetimes.\\
 
 The paper is organized as follows. In section 2 we present the geodesic deviation equation 
 expressed in terms of a suitably chosen vector basis on the geodesic and its first integrals 
 generated by the Killing vector fields . Section 3 deals with the problem of 
 maximal geodesics in de Sitter spacetime, in section 4 the same problem is studied in 
 anti--de Sitter space and in section 5 --- in Bertotti--Robinson spacetime. Conclusions 
 inferred from these cases are formulated in section 6.\\
 In a following paper we consider other spacetimes with high symmetry, first of all the 
 Reissner--Nordstr\"om one.
 
 \section{Equations for the Jacobi vector fields}
 Here we summarize for the reader's convenience, the formalism necessary for the search of locally 
 maximal timelike geodesics (cf. \cite{SG1}). 
 A Jacobi field on a given timelike geodesic $\gamma$ with a unit tangent vector field 
 $u^{\alpha}(s)$ is any vector field $Z^{\mu}(s)$ being a solution of the geodesic deviation equation on 
 $\gamma$, 
 \begin{equation}\label{n1}
 \frac{D^2}{ds^2}\,Z^{\mu}=R^{\mu}{}_{\alpha\beta\gamma}\,u^{\alpha}\,u^{\beta}\,Z^{\gamma},
\end{equation}
which is orthogonal to the geodesic, $Z^{\mu}\,u_{\mu}=0$. One replaces the second absolute derivative 
$D^2/ds^2$ by the ordinary one by expanding $Z^{\mu}$ in a basis consisting of three spacelike orthonormal 
vector fields $e_a{}^{\mu}(s)$, $a=1,2,3$ on $\gamma$, which are orthogonal to $\gamma$ and are parallelly 
transported along the geodesic, i.~e.  (the signature is $(+---)$)
\begin{equation}\label{n2}
 e_{a}{}^{\mu}\, e_{b\mu}=-\delta_{ab}, \qquad  e_{a}{}^{\mu}\,u_{\mu}=0, \qquad \frac{D}{ds}e_{a}{}^{\mu}=0.
\end{equation}   
In this basis $Z^{\mu}=\sum_a Z_a e_a{}^{\mu}$ and the covariant vector equation (1) is reduced to three scalar 
second order ODEs for the scalar coefficients $Z_a(s)$,
 \begin{equation}\label{n3}
 \frac{d^2}{ds^2}\,Z_a=-e_{a}{}^{\mu}\,R_{\mu\alpha\beta\gamma}\,u^{\alpha}\,u^{\beta}
 \sum_{b=1}^{3} Z_b\,e_{b}{}^{\gamma}.
\end{equation}  
A general Jacobi field depends on 6 integration constants appearing as a result of solving (3).\\

Any Killing vector field $K^{\mu}$ of the spacetime generates a first integral of eq. (1) of the form 
\cite{F1}
 \begin{equation}\label{n4}
 K_{\mu}\,\frac{D}{ds}Z^{\mu}-Z^{\mu}\,\frac{D}{ds}K_{\mu}=\textrm{const}.
\end{equation}  
By applying the derivative $D/ds$ to the function on the LHS of (4) one verifies that it is constant along 
the given geodesic. Also the integral of motion may be recast in terms of the scalars $Z_a$. To this end 
one introduces a spacetime tetrad $e_A{}^{\mu}$, $A=0,1,2,3$, along $\gamma$ consisting of the spacelike 
vectors $e_a{}^{\mu}(s)$ supplemented by $e_0{}^{\mu}\equiv u^{\mu}$. The tetrad is orthonormal,
\begin{equation}\label{n5}
 e_{A}{}^{\mu}\, e_{B\mu}=\eta_{AB}=\textrm{diag}(1,-1,-1,-1)
\end{equation}  
and parallelly transported along $\gamma$. Expanding $Z^{\mu}$ and $K^{\mu}$ in the tetrad, 
 $K^{\mu}=\sum_{A=0}^3 K_A\,e_A{}^{\mu}$ with the scalars $K_A$ defined by $K^{\mu}\,e_{A\mu}=\eta_{AA}
 K_A$ (no summation) and inserting them into (4) one gets
 \begin{equation}\label{n6}
 \sum_{a=1}^3 (Z_a\,\frac{dK_a}{ds}-\frac{dZ_a}{ds}\,K_a)=\textrm{const},
\end{equation}  
where $K_a=-K^{\mu}\,e_{a\mu}$. If the spacetime admits $n$ linearly independent 
Killing vector fields one gets $n$ integrals of motion (6). In a number of cases we find that 
some of these integrals are trivial, i.~e.~may be found without the use of the appropriate Killing fields and have already been employed at the very beginning of solving the relevant equations, whereas some other first integrals generated by independent Killing vectors turn out to be dependent. Nevertheless, in general the first integrals 
(6) are essential in the search for the Jacobi fields.
 
 \section{De Sitter spacetime}
We use the coordinates in which the spacetime is explicitly static \cite{SB}, \cite{BK}, \cite{GP},
 \begin{equation}\label{n7}
 ds^2=(1-H^2r^2)\,dt^2-(1-H^2r^2)^{-1}dr^2-r^2(d\theta^2+\sin^2\theta\,d\phi^2),
\end{equation}  
$t\in (-\infty,\infty)$, $0\leq r<1/H$, $t$ and $r$ have dimension of length. This chart covers only a part 
of the whole manifold; the spaces $t=$const are halves of three-spheres of constant radius $1/H$ and the 
spacetime is spherically symmetric. The Killing vector field $\partial/\partial t$ is timelike in the domain 
of the chart. The surface $r=1/H$ is a coordinate singularity and its part $t=-\infty$ is a past event 
horizon for an observer staying at $r=0$ whereas the part $t=+\infty$ is a future event horizon for the 
observer. The fact that the spacetime expands has a considerable influence on long journeys and on the 
possibility of communication during these journeys \cite{BMW}.\\

The static twin A remains at $r=r_0>0$, $\theta=\pi/2$, $\phi=\phi_0$ whereas the twin B moves on a circular 
orbit $r=r_0$ in the 2-surface $\theta=\pi/2$ with angular velocity $\omega=d\phi/dt$. The general 
geodesic equation for the radial coordinate,
 \begin{equation}\label{n8}
-\frac{\ddot{r}}{1-H^2r^2}-\frac{H^2r\dot{r}^2}{(1-H^2r^2)^2}+H^2r\dot{t}^2+r\dot{\theta}^2+r\dot{\phi}^2
\sin^2\theta=0, 
\end{equation}
where $\dot{f}\equiv df/ds$ throughout the paper, shows that circular worldlines $r=r_0>0$ cannot be 
geodesic curves, what is in concordance with the expansion of the spacetime. We assume for B that his 
nongeodesic motion is $\phi=\omega t+ \textrm{const}$ with $\omega=\textrm{const}>0$, then $\dot{\phi}=
\omega\dot{t}$.\\
All the motions take place in the 'plane' $\theta=\pi/2$, hence the universal integral of motion is
 \begin{equation}\label{n9}
g_{\alpha\beta}\,\dot{x}^{\alpha}\,\dot{x}^{\beta}= (1-H^2r^2)\,\dot{t}^2-
\frac{\dot{r}^2}{1-H^2r^2}-r^2\dot{\phi}^2=1. 
\end{equation} 
For the circular B's worldline the integral (9) yields
 \begin{equation}\label{n10}
\dot{t}^2=[1-r_0^2(H^2+\omega^2)]^{-1}
\end{equation} 
and this relation imposes an upper limit on $\omega$,
 \begin{equation}\label{n11}
\omega<\frac{1}{r_0}\sqrt{1-H^2r_0^2}.
\end{equation} 
Assuming that (11) holds and denoting 
 \begin{equation}\label{n12}
\beta\equiv[1-r_0^2(H^2+\omega^2)]^{-1/2} 
\end{equation} 
one gets for B,
\begin{equation}\label{n13} 
t(s)-t_0=\beta s \qquad \textrm{and} \qquad \phi-\phi_0=\omega(t-t_0).
\end{equation}
The coordinate time period $T$ of the B's circulation follows from $\phi(t_0+T)=\phi_0+2\pi$ and is 
$T=2\pi/\omega$, hence the proper time measured by B after making one full circle satisfies $T=\beta 
s_B(T)$ and is 
\begin{equation}\label{n14} 
s_B(T)=\frac{2\pi}{\beta\omega}.
\end{equation}
The length of the static A's worldline in the period $T$ is 
\begin{displaymath}
s_A(T)=\int_{t_0}^{t_0+T}\sqrt{1-H^2r_0^2}\,dt=\frac {2\pi}{\omega}\sqrt{1-H^2r_0^2}.
\end{displaymath}
Comparison of the lengths,
\begin{equation}\label{n15} 
\frac{s_A(T)}{s_B(T)}=\left(\frac{1-H^2r_0^2}{1-r_0^2(H^2+\omega^2)}\right)^{1/2}>1,
\end{equation}
confirms in this case the conjecture in \cite{ABK}, \cite{AB} that the moving faster twin is younger at 
the reunion than the static twin. In general, however, the conjecture is false.\\

The radially moving twin C has $\phi=\phi_0$ and the timelike Killing vector $K^{\alpha}=\delta^{\alpha}
_0$ (normalized to 1 at the coordinate singularity $r=0$) generates for his geodesic worldline the integral 
of energy $K^{\alpha}p_{\alpha}=E/c$, where $p^{\alpha}$ is the C's four-momentum. If the twin C has mass $m$ one defines a dimensionless constant of energy, $k\equiv E/(mc^2)$ and then 
\begin{equation}\label{n16} 
\dot{t}=\frac{k}{1-H^2r^2}.
\end{equation}
Inserting (16) into (8) and (9) one gets for a radial geodesic 
\begin{equation}\label{n17} 
(1-H^2r^2)\ddot{r}+H^2r(\dot{r}^2-k^2)=0
\end{equation}
\begin{equation}\label{n18} 
\textrm{and} \qquad \dot{r}^2=H^2r^2+k^2-1.                 
\end{equation}
The geodesic C emanates from the initial point $P_0(t=t_0, r=r_0>0, \phi=\phi_0)$. If $\dot{r}(t_0)=0$ 
then $k^2=1-H^2r_0^2$ and $\dot{r}\neq 0$ at later times implies $r>r_0$, i.~e.~the twin C will be 
eternally receding to infinity and will never return. We therefore assume that at $t_0$ it starts radially 
inwards with velocity $\dot{r}(t_0)=-u<0$. From (18), $u$ and $k$ are related by 
\begin{equation}\label{n19} 
k^2=u^2-H^2r_0^2+1.                 
\end{equation}
If $k=1$ the centre $r=0$ may be asymptotically reached for the proper time $s$ and coordinate time $t$ 
tending to infinity since $r=r_0\exp(-Hs)$.
 For $k>1$ the twin crosses the centre and moves outwards at $\phi=\phi_0+\pi$. We 
therefore assume $k<1$. Under this assumption at $t=t_m$ the twin reaches the smallest distance from 
the centre, $r=r_m$, where $\dot{r}(t_m)=0$ and the spacetime expansion makes it move outwards; at 
$t=t_1$ it returns to $r=r_0$ and this is spacetime point $P_1(t=t_1, r=r_0, \theta=\pi/2, \phi=
\phi_0)$. Clearly $t_1-t_m=t_m-t_0$ and the radial journey duration is $\Delta t=t_1-t_0=2(t_m-t_0)$. 
The lowest point $r_m$ is determined by $0=H^2r_m^2+k^2-1$, hence 
\begin{equation}\label{n20} 
r_m=\frac{1}{H}\sqrt{1-k^2}                
\end{equation}
and one sees from (19) that indeed $r_m<r_0$. The integral of energy on C is confined to the interval 
$\sqrt{1-H^2r_0^2}<k<1$.\\

The geodesic C from $P_0$ to $P_1$, i.~e.~on both the ingoing and outgoing segment, may be 
continuously parameterized as 
\begin{equation}\label{n21} 
r=r_m\,\cosh\eta, \qquad \ \eta\in [-\alpha, \alpha], \qquad \alpha>0.             
\end{equation}
By definition, $r(-\alpha)=r(\alpha)=r_0$, then $\cosh\alpha=r_0/r_m$. Inserting (21) into (18) and 
applying (20) one finds $d\eta/ds=H$. The length of C is counted from $P_0$, where $\eta=-\alpha$, 
then $\eta=Hs-\alpha$. The dependence $t(s)$ on the geodesic follows from (16),
\begin{equation}\label{n22} 
t(s)=\frac{k}{H}\int\frac{d\eta}{1-(1-k^2)\cosh^2\eta} = \frac{1}{H}\,
\textrm{ar tanh}\left[\frac{1}{k}\tanh(Hs-\alpha)\right]+\textrm{const}.            
\end{equation}
This function is well defined since the argument satisfies the inequality\\
 $|\frac{1}{k}\tanh\eta|<1$. 
In fact, $|\tanh\eta|\leq\tanh\alpha$ and one infers from (20) that
\begin{displaymath}
1-\frac{1}{k^2}\tanh^2\alpha=1-\frac{r_0^2-r_m^2}{k^2r_0^2}=\frac{1-k^2}{k^2H^2r_0^2}(1-H^2r_0^2)>0.
\end{displaymath}
Inverting the function(21) one gets the length of C from $r_0$ to $r_m$,
\begin{equation}\label{n23} 
s(r_0,r_m)=\frac{\alpha}{H} = \frac{1}{H}\,\textrm{ar cosh}\frac{r_0}{r_m}          
\end{equation}
and the length of the geodesic from $P_0$ to $P_1$ expressed in terms of $r_0$ and $k$ is 
\begin{equation}\label{n24} 
s_C=2s(r_0,r_m)=-\frac{1}{H}\ln(1-k^2)+ \frac{2}{H}\,\ln\left[Hr_0+\sqrt{H^2r_0^2+k^2-1}\right].          
\end{equation}
Applying properties of hyperbolic functions one arrives at
 \begin{eqnarray}\label{n25}
t(\eta)-t_0 & = & \frac{1}{H}\,\textrm{ar tanh}\left(\frac{1}{k}\tanh\eta\right)+
\frac{1}{H}\,\textrm{ar tanh}\left(\frac{1}{k}\tanh\alpha\right)
\nonumber\\
& = & \frac{1}{H}\,\textrm{ar tanh}\left(\frac{k(\tanh\eta+\tanh\alpha)}{k^2+\tanh\eta\,\tanh\alpha}
\right)
\end{eqnarray} 
and the coordinate time of the flight from $P_0$ to $P_1$ is
\begin{equation}\label{n26} 
\Delta t=2(t_m-t_0)=\frac{2}{H}\,\textrm{ar tanh}\left(\frac{1}{k}\tanh\alpha\right)= 
\frac{1}{H}\,\ln\frac{k+\tanh\alpha}{k-\tanh\alpha},         
\end{equation} 
\begin{equation}\label{n27}
\textrm{where} \qquad \tanh\alpha= \frac{1}{Hr_0}\sqrt{H^2r_0^2+k^2-1}.         
\end{equation} 
The three twins depart from $P_0$, yet one cannot assume that they will meet together at $P_1$. A 
and C will meet at this event whereas in general B and C cannot. We shall discuss the latter problem 
below, now we compare the proper times of A and C. The length of A's worldline between $P_0$ and 
$P_1$ is $s_A(\Delta t)=(1-H^2r_0^2)^{1/2}\Delta t$ and 
\begin{equation}\label{n28}
\frac{s_C}{s_A(\Delta t)}=(1-H^2r_0^2)^{-1/2}[2\ln(Hr_0+Q)-\ln(1-k^2)]\left(\ln\frac{kHr_0+Q}
{kHr_0-Q}\right)^{-1},
\end{equation} 
where $Q\equiv\sqrt{H^2r_0^2+k^2-1}$. A number of numerical examples confirms the expectation that 
always $s_C>s_A(\Delta t)$, e.~g.~for $Hr_0=0,99$ and $k=0,5$ the ratio is $s_C/s_A=2,2654$.\\

The twins B and C will meet at $P_1$ if B's angular velocity is $\omega=2\pi/\Delta t$ for $\Delta t$ 
given in (26) and if $\omega$ is smaller than the upper limit (11) what amounts to 
\begin{equation}\label{n29}
\ln\frac{kHr_0+Q}{kHr_0-Q}>\frac{2\pi Hr_0}{\sqrt{1-H^2r_0^2}};
\end{equation} 
for fixed $r_0$ it is a restriction of the C's energy $k$. In the example above, $Hr_0=0,99$ and $k=0,5$ 
the twin B cannot meet C after one revolution. The two twins will meet at $P_1$ if the energy $k$ 
is sufficiently close to 1: if $k=1-\varepsilon$ with $0<\varepsilon\ll 1$ then
\begin{displaymath}
\ln\frac{kHr_0+Q}{kHr_0-Q}\approx \ln\left(\frac{2Hr_0}{\varepsilon}-\frac{1+H^2r_0^2}{1-H^2r_0^2}
\right)
\end{displaymath}
and inequality (29) holds. 

 \subsection{Jacobi fields on the timelike geodesics in de Sitter space}
Timelike geodesics in de Sitter spacetime have no conjugate points. In fact, according to Proposition 
4.4.2 in \cite{HE} (cited as Proposition 4 in \cite{S}) the necessary conditions are not satisfied: 
whereas the condition $R_{\mu\alpha\nu\beta}\,u^{\alpha}u^{\beta}=\frac{R}{12}(g_{\mu\nu}-u_{\mu}u_{\nu})
\neq 0$ holds for any timelike geodesic, the expression $R_{\alpha\beta}u^{\alpha}u^{\beta}=R/4=-3H^2$ 
is always negative. For completeness we also investigate the existence of conjugate points on null geodesics. 
According to Proposition 4.4.5 in \cite{HE} the null tangent vector $k^{\alpha}$ should satisfy two 
necessary conditions: $R_{\alpha\beta}k^{\alpha}k^{\beta}\geq 0$ and it holds since the scalar is zero and 
$k^{\mu}k^{\nu}k_{[\alpha} R_{\beta]\mu\nu[\lambda}k_{\sigma]}\neq 0$ at a point and this does not hold 
since the tensor vanishes identically. Hence de Sitter spacetime has no future (nor past) nonspacelike 
conjugate points and because the other two assumptions of Theorem 11.16 in \cite{BEE} (cited as Theorem 6 
in \cite{SG1}) concerning the spacetime hold, one concludes that each timelike geodesic is the 
unique longest (i.~e.~maximal) curve connecting its endpoints (it contains no cut points).  
In other terms the endpoint $P_1$ is in a convex normal neighbourhood of an arbitrarily chosen 
point $P_0$ (assuming $P_0\prec\prec P_1$).\\

We now determine a general Jacobi field on any timelike geodesic. 
In de Sitter space the geodesic deviation equation (3) for the 
scalar coefficients $Z_a(s)$ is the same for all geodesic curves, whether radial or not and for 
any choice of the spacelike triads on them (providing they satisfy (2)) and has the form 
\begin{equation}\label{n30}
\frac{d^2Z_a}{ds^2}-H^2\,Z_a=0.
\end{equation} 
The generic solution is 
\begin{equation}\label{n31}
Z_a=C_a\,e^{Hs}+C'_a\,e^{-Hs}.
\end{equation} 
Then the general deviation vector field on arbitrary timelike geodesic vanishing at the initial point 
$P_0(s=0)$ is
\begin{equation}\label{n32}
Z^{\mu}(s)=\sum_{a=1}^{3}C_a\,e_a{}^{\mu}(s)\sinh Hs
\end{equation} 
and it is clear that there are no conjugate points on it since the neighbouring geodesics 
exponentially diverge. The gravitation in this spacetime is repulsive.\\
Finally we explicitly determine the general Jacobi field on a radial timelike geodesic. For generality 
we consider the geodesic followed by the twin C, i.~e.~consisting of the ingoing segment and the 
outgoing one. The vector tangent to C is, from (16), (18) and (21),
\begin{eqnarray}\label{n33}
u^{\alpha} & = & \left[\frac{k}{1-H^2r^2}, \varepsilon\sqrt{H^2r^2+k^2-1}, 0, 0\right]=
\nonumber\\
& = & \left[\frac{k}{1-H^2r_m^2\cosh^2\eta}, \sqrt{1-k^2}\sinh\eta, 0, 0\right]
\end{eqnarray} 
with $\eta=Hs-\alpha$ and $\varepsilon=-1$ on the ingoing segment, $-\alpha\leq\eta<0$ and 
$\varepsilon=+1$ on the outgoing piece, $0<\eta\leq\alpha$. The basis vector triad on C satisfying (2) 
is chosen as
 \begin{eqnarray}\label{n34}
e_1{}^{\mu} & = & \left[\frac{\sqrt{1-k^2}}{1-H^2r_m^2\cosh^2\eta}\,\sinh\eta, k, 0, 0\right],
\nonumber\\
e_2{}^{\mu} & = & \left[0,0,\frac{1}{r_m\cosh\eta},0\right], \qquad  
e_3{}^{\mu} = \left[0,0,0,\frac{1}{r_m\cosh\eta}\right],   
\end{eqnarray} 
$-\alpha\leq\eta\leq +\alpha$. One inserts (34) into (32). 

\section{Anti-de Sitter spacetime}
Actually we consider the covering anti-de Sitter (CAdS) spacetime and use the chart covering the entire 
manifold and exhibiting its static nature; the radial coordinate is suitably chosen to our purposes 
\cite{HE}, \cite{GP}, \cite{G}
\begin{equation}\label{n35}
ds^2=\frac{r^2+a^2}{a^2}\,dt^2-\frac{a^2}{r^2+a^2}\,dr^2-r^2(d\theta^2+\sin^2\theta\,d\phi^2),
\end{equation} 
where $t\in(-\infty,+\infty)$, $r\in[0,\infty)$, $t$, $r$ and $a$ have dimension of length. Hypersurfaces 
of simultaneity $t=$const are the hyperbolic Lobatchevski spaces $H^3$. There are no horizons and 
$r=0$ is a coordinate singularity.\\
The non-geodesic static twin A stays at $r=r_0>0$ and $\phi=\phi_0$ and in a coordinate time interval 
$T$ his worldline has length 
\begin{equation}\label{n36}
s_A(T)=\sqrt{\left(\frac{r_0}{a}\right)^2+1}\,\,T.
\end{equation}
Any geodesic motion takes place in the 2-surface $\theta=\pi/2$. For a timelike geodesic one has the 
integral of energy $E$ generated by the timelike Killing field $K^{\alpha}=\delta^{\alpha}_0$  
(normalized to 1 at $r=0$), $K^{\alpha}p_{\alpha}=E/c$, then $k\equiv E/(mc^2)$ and 
\begin{equation}\label{n37}
\dot{t}\equiv \frac{dt}{ds}=\frac{a^2k}{r^2+a^2}.
\end{equation}
The rotational Killing field $\partial/\partial \phi$ with components $\xi^{\alpha}=
\delta^{\alpha}_3$ is normalized as in Minkowski space and gives rise to conserved angular momentum 
$J=-\xi^{\alpha}p_{\alpha}$. Introducing a dimensionless angular momentum $h$ by $ah=J/(mc)$ one 
gets
\begin{equation}\label{n38}
\dot{\phi}= \frac{ah}{r^2}.
\end{equation} 
The geodesic equation for the radial coordinate, upon employing (37) and (38) is
\begin{equation}\label{n39}
(r^2+a^2)\ddot{r}-r\dot{r}^2+k^2r-\frac{h^2}{r^3}(r^2+a^2)^2=0
\end{equation}
and the universal integral of motion $g_{\alpha\beta}\,\dot{x}^{\alpha}\dot{x}^{\beta}=1$ reads 
\begin{equation}\label{n40}
\dot{r}^2=k^2-\frac{r^2+a^2}{a^2}-\frac{h^2}{r^2}(r^2+a^2).
\end{equation}
A circular geodesic at any $r=r_0>0$ does exist and is stable \cite{SG1} and the constants of motion 
are in this case determined from (39) and (40) by $a$ and $r_0$ as
\begin{equation}\label{n41}
k=\frac{r_0^2+a^2}{a^2} \qquad \textrm{and} \qquad h=\frac{r_0^2}{a^2}.
\end{equation}
As the starting point for the three worldlines we choose $P_0(t_0=0, r=r_0>0, \theta=\pi/2, \phi=
\phi_0)$ then the circular geodesic B has the form 
\begin{equation}\label{n42}
t=s, \qquad \textrm{and} \qquad \phi-\phi_0=\frac{s}{a}=\frac{t}{a},
\end{equation} 
what implies constant angular velocity 
$\omega=d\phi/dt=1/a$. The period of one revolution is $T=2\pi a$ and the length of geodesic B 
corresponding to one circle, i.~e.~between $t=0$ and $t=T$ is $s_B=T=2\pi a$. Both $T$ and $s_B$ are 
the same for all circular geodesic curves independently of the radius $r_0$; it is a trace of the 
original anti-de Sitter spacetime where all timelike curves are closed with a period $2\pi a$. 
Clearly $s_A(2\pi a)>s_B$.\\

The third twin C moves on a radial geodesic $h=0$ and the equations describing it are reduced to 
\begin{equation}\label{n43}
(r^2+a^2)\ddot{r}-r\dot{r}^2+k^2r=0
\end{equation}
\begin{equation}\label{n44}
\textrm{and} \qquad \dot{r}^2=k^2-\frac{r^2+a^2}{a^2}.
\end{equation}
Let at $P_0$ the twin C be initially at rest, $\dot{r}(t=0)=0$, then its energy $k$ is 
given by $k^2=r_0^2/a^2+1$ 
and its acceleration is directed downwards, $\ddot{r}(0)=-r_0/a^2<0$, implying falling down. This 
shows that gravitation in CAdS is attractive: a body left at rest falls radially to the centre,  
reaches the centre $r=0$ and flies away in the opposite direction $\phi=\phi_0+\pi$. 
From (44) it follows $\dot{r}^2=(r_0^2-r^2)/a^2$ and this implies $r\leq r_0$. 
At the antipodal point $r=r_0$, $\phi=\phi_0+\pi$ (denoted below as $P_3$) 
there is again $\dot{r}=0$ and $\ddot{r}=-r_0/a^2$ and the body falls down back and returns to the 
starting point at the space, $r=r_0$, $\phi=\phi_0$. We therefore assume that C moves as in 
Schwarzschild spacetime: it radially flies away with initial velocity $\dot{r}(0)=u>0$, reaches 
a maximum height $r=r_M$ at $t=t_M$ and falls down back to $r_0$ at the event 
$P_1(t=t_1, r=r_0, \phi=\phi_0)$ where $t_1=2t_M$. The quantities $r_0$, $k$ and $u$ are now related by
\begin{equation}\label{n45}
u^2=\frac{1}{a^2}(a^2k^2-r_0^2-a^2)
\end{equation}
and the highest point of the trajectory is
\begin{equation}\label{n46}
r_M^2=a^2(k^2-1).
\end{equation}
The condition $r_M>r_0>0$ implies 
\begin{equation}\label{n47}
k^2>\frac{r_0^2+a^2}{a^2}.
\end{equation}
One sees from (46) that $r_M<\infty$ for $k<\infty$ what implies that a massive particle with finite 
energy cannot escape to the spatial infinity $r=\infty$ (\cite{AIS}, \cite{GP} par. 5.2). This 
property of CAdS is in marked contrast to Schwarzschild spacetime, where the corresponding 
relationship is \cite{S} $r_M=2M/(1-k^2)$ and $r_M$ tends to infinity for $k\rightarrow 1$ from 
below.\\

On the segment $P_0P_1$ (and possibly outside it) the radial geodesic C is conveniently parameterized 
by an angle $\eta$,
\begin{equation}\label{n48}
r(\eta)=r_M\,\cos^2\eta\equiv\frac{1}{2}r_M (\cos2\eta+1)
\end{equation} 
in the interval $-\alpha/2\leq\eta\leq+\alpha/2$. Then $r_M=r(0)$ and the endpoints $P_0$ and $P_1$ 
correspond to $r_0=r(-\alpha/2)=r(+\alpha/2)$ and the boundary angle $\alpha$ is determined by 
\begin{equation}\label{n49}
\cos\alpha=\frac{2r_0}{r_M}-1
\end{equation} 
and $\cos\alpha$ is bounded from above by 
\begin{displaymath}
\cos\alpha=\frac{1}{r_M}(2r_0-r_M)=\frac{1}{r_M}(2r_0-a\sqrt{k^2-1})<\frac{1}{r_M}\left(2r_0-
a\sqrt{\frac{r_0^2}{a^2}}\right)=\frac{r_0}{r_M},
\end{displaymath}
so that $\arccos\frac{r_0}{r_M}<\alpha<\pi$. \\
The geodesic C consists of the outgoing segment, $\eta\in[-\alpha/2, 0)$ and the ingoing one, 
$\eta\in(0,+\alpha/2]$. The time component of the vector $u^{\alpha}$ tangent to C is given in  (37); 
its radial component is from (44) $\dot{r}=\varepsilon(k^2-r^2/a^2-1)^{1/2}$, where $\varepsilon=
+1$ on the outgoing segment and $\varepsilon=-1$ on the ingoing one. Applying (48) in $\dot{r}$ 
and noticing that $\varepsilon(\sin^2\eta)^{1/2}=-\sin\eta$ both on the outgoing and on the 
ingoing segment, one finally arrives at
\begin{equation}\label{n50}
u^{\alpha}\equiv \dot{x}^{\alpha}=\left[\frac{k}{(k^2-1)\cos^4\eta+1}, -(k^2-1)^{1/2}\sin\eta\,
(1+\cos^2\eta)^{1/2}, 0, 0\right]
\end{equation}  
valid along the whole geodesic line. The derivative $ds/d\eta$ along C may be found from the 
expression for $ds^2$ on C by inserting $dr/d\eta=-r_M\sin2\eta$ and $dt/d\eta=dt/ds\cdot 
ds/d\eta$ and applying (37); the resulting equation is solved by 
\begin{equation}\label{n51}
\frac{ds}{d\eta}=\frac{2a|\cos\eta|}{(1+\cos^2\eta)^{1/2}}.
\end{equation}  
If $\sin\eta$ is a monotonic function in an interval $\eta_1<\eta<\eta_2$ the length of the 
corresponding geodesic arc is 
\begin{equation}\label{n52}
s(\eta_1,\eta_2)=\int^{\eta_2}_{\eta_1}ds=2\sigma a\left[\arcsin\left(
\frac{\sin\eta_2}{\sqrt{2}}\right)-\arcsin\left(\frac{\sin\eta_1}{\sqrt{2}}\right)\right],
\end{equation}  
where $\sigma=+1$ if $\cos\eta>0$ in the interval and $\sigma=-1$ for $\cos\eta<0$. Then the length 
of the geodesic C from $P_0$ to $P_1$ is 
\begin{equation}\label{n53}
s_C=s(-\frac{\alpha}{2},+\frac{\alpha}{2})=2s(-\frac{\alpha}{2},0)=2s_M=4a
\arcsin\left(\frac{\sin\frac{\alpha}{2}}{\sqrt{2}}\right)=2a\arccos\left(\frac{r_0}{r_M}\right),
\end{equation}  
where $r_0/r_M=r_0/(a\sqrt{k^2-1})$. \\
To calculate the coordinate time interval corresponding to a given segment of the radial geodesic 
one writes $dt/d\eta=dt/ds\cdot ds/d\eta$ and inserts (37), (48) and (51), then 
\begin{equation}\label{n54}
\frac{dt}{d\eta}=-\frac{8\varepsilon a k\sin2\eta}{(k^2-1)(\cos2\eta+1)^2+4}[4-(\cos2\eta+1)^2]^
{-1/2},
\end{equation} 
here as above $\varepsilon=+1$ on the outgoing segment ($\sin2\eta<0$) and $\varepsilon=-1$ on 
the outgoing segment ($\sin2\eta>0$). One then gets
\begin{equation}\label{n55}
\Delta t(\eta_1,\eta_2)=\int^{\eta_2}_{\eta_1}\frac{dt}{d\eta}\, d\eta=
\varepsilon a[\arctan(f(\eta_2))-\arctan(f(\eta_1))],
\end{equation}  
\begin{displaymath}
\textrm{here} \qquad f(\eta)\equiv \frac{k(\cos2\eta+1)}{\sqrt{4-(\cos2\eta+1)^2}}.
\end{displaymath}
The time of the flight from $P_0$ to $P_1$ is 
\begin{eqnarray}\label{n56}
t_1 & = & \Delta t(-\frac{\alpha}{2},+\frac{\alpha}{2})=2t_M=\pi a-2a
\arctan\left(\frac{kr_0}{\sqrt{r_M^2-r_0^2}}\right)
\nonumber\\
& = & 2a\,\arccos\left(\frac{kr_0}{(k^2-1)^{1/2}(a^2+r_0^2)^{1/2}}\right).
\end{eqnarray} 
For $r_0\rightarrow\infty$ one has $k^2\rightarrow(r_0/a)^2\rightarrow\infty$ and 
$r_0/r_M\rightarrow1$, thence $s_C\rightarrow 0$ and $t_1\rightarrow 0$. CAdS spacetime has the 
peculiar feature that both the length of any radial timelike geodesic (emanating from $r_0<\infty$ and 
consisting of the outgoing segment and ingoing one returning to $r_0$) and the coordinate time of 
flight on it are bounded from above: $t_1<\pi a$ and $s_C<\pi a$; correspondingly a radial 
geodesic consisting of one outgoing or ingoing segment from $r_0$ to $r_M$ has $s(r_0,r_M)<
\pi a/2$. These upper limits correspond to $r_0\rightarrow 0$ independently of $r_M$. Yet the 
spatial distance (the length of a radial spacelike geodesic at $t=$const) from any finite $r_0$ to 
$r=\infty$ is infinite. Furthermore, it has been shown that in the 2-surface $(t,r)$ there exist 
points $p$ and $q$ which are chronologically related ($q$ lies inside the future null cone of $p$) and 
such that there is no timelike (and necessarily radial) geodesic joining them \cite{CM}, \cite{Pen}, 
\cite{AIS}, \cite{BEE} (Chap. 6).\\

 The twins B and C start from the same place, yet they will not meet again after making one 
 circle and one radial flight, respectively, since $T=2\pi a>t_1$. There are, however, two 
 cases in which they can reunion.\\
 1. Let C make a number of radial flights back and forth in such a way that at $r=r_0$ it  
 bounces, i.~e.~rapidly alters its radial velocity from $-u$ to $+u$. Its worldline consists 
 of a number of smooth geodesic segments which are non-smoothly joined at $r_0$, it forms a 
 broken geodesic. Moreover let the energy $k$ of C be suitably tuned (for fixed $r_0$) so that 
\begin{displaymath}
\frac{kr_0}{(k^2-1)^{1/2}(a^2+r_0^2)^{1/2}}=\cos\left(\frac{m}{n}\pi\right),
\end{displaymath}     
where $m$ and $n$ are positive integers and $m<n$. Then duration of $n$ consecutive radial flights 
is $nt_1=2m\pi a=mT$, the duration of orbiting $m$ full circles by B. When they meet at $r_0$ at 
$t=nt_1$ their proper times are $ns_C$ and $ms_B$ and their difference is 
\begin{displaymath}
ns_C-ms_B=2na\left(\arccos\frac{r_0}{r_M}-\frac{m}{n}\pi\right).
\end{displaymath}
We compare the two angles by taking the ratio of their cosines,
\begin{displaymath}
\cos\left(\arccos\frac{r_0}{r_M}\right)\left[\cos\left(\frac{m}{n}\pi\right)\right]^{-1}=
\left(\frac{r_0^2+a^2}{a^2k^2}\right)^{1/2}<1
\end{displaymath}  
according to (47) and one infers that $m\pi/n$ is the smaller angle and $ns_C-ms_B>0$.\\

2. A physically more interesting opportunity is that the falling down twin C is not stopped at 
$r_0$ and is allowed to freely move farther. Then the whole sequence of events is following:\\
C starts from $r_0$ at $P_0(t=0,\eta=-\alpha/2)$, reaches $r_M$ at $t_M$ and falls down, comes 
back to $r_0$ at $P_1(t=t_1, \eta=+\alpha/2)$, arrives at the centre at $P_2(t=t_2, \eta=\pi/2)$, 
crosses it and radially flies upwards at $\phi=\phi_0+\pi$, passes by the opposite point $r_0$ 
at $P_3(t=t_3, \eta=\pi-\alpha/2)$, gets to the highest point $r_M$ at $P_4(t=t_4,\eta=\pi)$ 
and turns downwards, falls down to $r_0$ at $P_5(t=t_5,\eta=\pi+\alpha/2)$, comes back to the 
centre $r=0$ at $P_6(t=t_6,\eta=3\pi/2)$ and finally returns to the starting place $r=r_0$ and 
$\phi=\phi_0$ at $P_7(t=t_7, \eta=2\pi-\alpha/2)$ with the initial velocity $+u$.\\
The staticity of the metric gives rise to the symmetry properties of the segments of this 
worldline. Employing (52), (55) and (56) one gets the coordinate time intervals:
\begin{equation}\label{n57}
t_2-t_1=t_3-t_2=t_6-t_5=t_7-t_6=a\,\arctan\left(\frac{kr_0}{\sqrt{r_M^2-r_0^2}}\right),
\end{equation} 
\begin{equation}\label{n58}
t_4-t_3=t_5-t_4=t_M
\end{equation} 
and the lengths of the corresponding geodesic segments,
\begin{equation}\label{n59}
s(P_1P_2)=s(P_2P_3)=s(P_5P_6)=s(P_6P_7)=\frac{\pi}{2}a-a\,\arccos\left(\frac{r_0}{r_M}\right),
\end{equation} 
\begin{equation}\label{n60}
s(P_0P_1)=s_C=2s_M, \qquad 
s(P_3P_4)=s(P_4P_5)=s_M.
\end{equation} 
Adding these seven segments one finds the time duration of the full cycle and the length of the 
geodesic,
\begin{equation}\label{n61}
\Delta t(-\frac{\alpha}{2},2\pi-\frac{\alpha}{2})=t_7=2\pi a, \qquad 
s(P_0P_7)=2\pi a.
\end{equation}
Both the circular geodesic B and the radial geodesic C emanating from any point $r_0>0$ reconverge at 
$P_7$ at the coordinate time $T=t_7=2\pi a$ having the same length $2\pi a$ and then at 
$t=4\pi a, 6\pi a,\ldots$ (\cite{AIS}, \cite{G}) independently of the initial velocity of C. 
Once again we emphasize that this is a trace of the original anti-de Sitter spacetime which is 
periodic in time, i.~e.~events $(t=0,r,\theta,\phi)$ and $(t=2\pi a,r,\theta,\phi)$ are identified. 
What is even more interesting here is that 
all timelike geodesics emanating from $P_0$ actually intersect at $P_3(r=r_0, \phi=\phi_0+\pi)$ 
which is spatially the antipodal point (with respect to the centre) to $P_0$. In fact, from (57) 
and (56) one gets $t(P_3)=t_3=\pi a$ and from (60), (59) and (53) it follows $s(P_0P_3)=\pi a$. 
By continuity it follows that the same holds for the radial geodesic which falls down from rest 
at $P_0$, i.~e.~$\dot{r}(0)=0$. The circular timelike geodesic B also intersects all the radial 
geodesics at $P_3$ since from (42) one finds that for $t=t_3=\pi a$ its angular coordinate is 
$\phi=\phi_0+\pi$ and $s_B(\pi)=\pi a$. Thus we have shown analytically that the circular and 
all radial geodesics (which cross $r=0$) emanating from $P_0$ do meet again after $\Delta t=
\pi a$ at the antipodal point $P_3$ and all have the same length. $P_3$ is the future cut 
point of $P_0$ lying on all radial and circular timelike geodesics.\\                                   

After one radial travel upwards and downwards the twin C meets A at $P_1$ and they compare their 
proper times,
\begin{equation}\label{n62}
\frac{s_C}{s_A(t_1)}=\frac{1}{\sqrt{(\frac{r_0}{a})^2+1}}\,\frac{\arccos\left[\frac{r_0}
{a(k^2-1)^{1/2}}\right]}{\arccos\left[kr_0(k^2-1)^{-1/2}(r_0^2+a^2)^{-1/2}\right]}.
\end{equation}
It is not easy to analytically prove that $s_C>s_A$. We do it numerically and in table 1 we give 
the ratio $s_C/s_A$ for $r_0=a$ and 5 values of $k$; it follows from (47) that $k^2>2$.
 \begin{table}
\caption{The ratio $s_C/s_A$ for $r_0=a$ as a function of energy $k$.}
\begin{tabular}{ll}
\hline\noalign{\smallskip}
 $k$  &    $s_C/s_A$ \\
 \noalign{\smallskip}\hline\noalign{\smallskip}
 $\frac{10}{9}\sqrt{2}$  &   1,0331 \\
                    2    &  1,0975 \\
 $\frac{5}{3}\sqrt{2}$   &  1,1351 \\
         $2\sqrt{2}$     &  1,1722 \\
        $10\sqrt{2}$     &  1,3547 \\
 \noalign{\smallskip}\hline 
\end{tabular}
\end{table} 
For $r_M\gg r_0$, i.~e.~for $k\rightarrow\infty$ the ratio $s_C/s_A$ tends to $\sqrt{2}$.

\subsection{Conjugate points on timelike geodesics}
In CAdS spacetime the necessary conditions $R_{\mu\alpha\nu\beta}\,u^{\alpha}u^{\beta}=\frac{1}{a^2}
(g_{\mu\nu}-u_{\mu}u_{\nu})\neq 0$ and $R_{\alpha\beta}u^{\alpha}u^{\beta}=3/a^2>0$ imply that 
each timelike geodesic contains conjugate points provided it is sufficiently extended. To determine 
conjugate points on a given geodesic one does not need to know the Jacobi vector fields 
associated with this geodesic. This is due to the fact that in CAdS the right-hand side of 
eq. (3) is universal (as is the case of de Sitter spacetime): is independent of the 
form of the tangent vector $u^{\alpha}$ and the spacelike basis fields $e_b{}^{\mu}(s)$, $b=1,
2,3$ and is determined solely by the curvature tensor and relations (2). This means in turn 
that the Jacobi scalars $Z_b(s)$ are universal and for all timelike geodesics they satisfy 
\begin{equation}\label{n63}
\frac{d^2}{ds^2}Z_b+\frac{1}{a^2}\,Z_b=0
\end{equation}
with the general solution (the change of sign in eq. (30) results in replacing exponential functions 
by trigonometric ones)
\begin{equation}\label{n64}
Z_b(s)=C_{b1}\sin\frac{s}{a}+C_{b2}\cos\frac{s}{a},
\end{equation}
$C_{b1}$, $C_{b2}$ arbitrary constants. Let $P_0$ be any point on the given geodesic chosen 
as the initial point ($s=0$), one seeks for points conjugate to $P_0$. The triad components 
$Z_b$ must vanish at $P_0$ and under this condition they reduce to $Z_b(s)=
C_{b1}\sin\frac{s}{a}$. These scalars have an infinite sequence of zeros at $s_n=n\pi a$, 
$n=1,2,\ldots$. In other words each point on each timelike geodesic has a point conjugate 
to it at a geodesic distance $\Delta s=\pi a$ and the sequence of conjugate points is infinite. 
Each of the three Jacobi vector fields corresponding to $Z_b(s)$ generates the sequence. \\

In the case of the circular curve B the first conjugate point to $P_0$ is that lying 
in the middle of the geodesic segment corresponding to one full revolution, i.~e.~half way 
between $P_0$ and $P_7$; it coincides with the first future cut point $P_3$. Further 
conjugate points (the second one is $P_7$) at $s_n=n\pi a$ are identical with the 
subsequent future cut points. In \cite{SG1} it was shown that if a static spherically symmetric 
spacetime admits stable timelike circular geodesics, then in general there exist on them 
three distinct infinite sequences of conjugate points. Due to the maximal symmetry of CAdS 
space one expects that these sequences should coincide and in fact, applying appropriate 
formulae from \cite{SG1} one easily checks that this is the case.\\

The same holds for radial timelike geodesics which oscillate between spatially antipodal 
highest points $r=r_M$: the subsequent conjugate points to $P_0$ coincide with their 
future cut points. Yet the radial geodesic which does not cross the centre $r=0$ is free 
of conjugate points. CAdS spacetime is not globally hyperbolic and the theorems quoted as 
Theorems 2 to 6 in  section 3 of \cite{SG1} do not apply. By symmetry considerations 
one expects that the geodesic C has no future cut points of $P_0$ earlier than $P_3$ and 
is maximal on the segment $P_0P_1$ whose length is $s_C<\pi a$ and this implies that 
$s_C>s_A(t_1)$.

\subsection{Jacobi fields on timelike radial and circular geodesics}
 According to (64) for each timelike geodesic the general Jacobi field has the same form
\begin{equation}\label{n65}
Z^{\alpha}(s)=\sum_{b=1}^{3}(C_{b1}\sin\frac{s}{a}+C_{b2}\cos\frac{s}{a})\,e_b{}^{\alpha},
\end{equation}
only the basis vectors $e_b{}^{\alpha}(s)$ depend on the given curve.\\
         
The basis of spacelike vector fields on the radial geodesic C which satisfy (2) may be 
chosen as
\begin{eqnarray}\label{n66}
e_1{}^{\alpha} & = & \left[\frac{\varepsilon a}{r^2+a^2}\left(a^2(k^2-1)-r^2\right)^{1/2}, 
k, 0,0\right],  
\nonumber\\
e_2{}^{\alpha} & = & \left[0,0,\frac{1}{r},0\right], \qquad 
e_3{}^{\alpha}=\left[0,0,0, \frac{1}{r}\right],
\end{eqnarray}
where $\varepsilon=+1$ on the outgoing segment $(-\alpha/2\leq\eta<0$) and 
$\varepsilon=-1$ on the ingoing one ($0<\eta\leq\pi/2$) and (48) holds. The component 
$e_1{}^0$ is continuous at $r=r_M$ where it changes its sign since it vanishes there. The Jacobi 
vector field $Z_2e_2{}^{\alpha}+Z_3e_3{}^{\alpha}$ connecting C to a nearby geodesic is directed off 
the 2--surface $t-r$ given by $\theta=\pi/2$ and the field $Z_1e_1{}^{\alpha}$ lies in the surface.  \\

In \cite{SG1} it was shown that the basis triad of vectors satisfying (2) on the circular geodesic B 
has a universal form common to all SSS spacetimes, depending on four constants, whose values in turn 
depend, for the given value of $r_0$, on the metric functions $g_{00}$ and $g_{11}$. In the present 
case these read
\begin{eqnarray}\label{n67}
e_1{}^{\alpha} & = & \left[-\frac{r_0}{(r_0^2+a^2)^{1/2}}\sin\frac{s}{a}, \frac{1}{a}
(r_0^2+a^2)^{1/2}\cos\frac{s}{a}, 0, -\frac{1}{ar_0}(r_0^2+a^2)^{1/2}\sin\frac{s}{a}
\right], 
\nonumber\\
 e_2{}^{\alpha} & = & \left[0,0,\frac{1}{r_0},0\right], \qquad
e_3{}^{\alpha} =-a\,\frac{d}{ds}e_1{}^{\alpha}.
\end{eqnarray}
All the three Jacobi fields are directed off the 2--surface $t-\phi$ where B lies.

\section{Bertotti--Robinson spacetime}
This spacetime, first discovered by T. Levi--Civita (1917), was independently rediscovered by 
Bertotti \cite{Be} and Robinson \cite{Ro}. The spacetime is homogeneous and spatially 
homogeneous, static, spherically symmetric (it admits a 6-dimensional isometry group) and 
conformally flat and it is a unique spacetime generated by a homogeneous non-null 
electromagnetic field; it also arises as a near--horizon limit of the non-extremal 
Reissner--Nordstr\"om black hole \cite{OT}. It is geodesically complete and it is 
conjectured that this spacetime and the Melvin solution are the only geodesically complete 
static Einstein--Maxwell spacetimes \cite{GG}; topologically it is $\textrm{AdS}_2\times S^2$, 
thus it is not globally hyperbolic (for a fuller description see \cite{SKMHH} par. 12.3 
and \cite{GP} par. 7.1). We investigate it in the chart
\begin{equation}\label{n68}
ds^2=a^2(\sinh^2x\,dt^2-dx^2-d\theta^2-\sin^2\theta\,d\phi^2),
\end{equation}                                                
where $a$ has dimension of length, $t\in(-\infty,+\infty)$, $x\in(0,\infty)$, all the 
coordinates are dimensionless and $x=0$ is a coordinate singularity. The timelike Killing 
vector chosen as $K^{\alpha}=\frac{1}{a}\,\delta^{\alpha}_0$ becomes null on the 
hypersurface $x=0$ which has topology $\mbox{\boldmath R}^{1}\times S^2$. The conserved energy 
for a geodesic motion generated by $K^{\alpha}$ is as usual $K^{\alpha}p_{\alpha}=E/c$ and the 
standard definition $k=E/(mc^2)$ yields the first integral, 
\begin{equation}\label{n69}
\dot{t}\equiv \frac{dt}{ds}=\frac{k}{a\sinh^2x},
\end{equation}                                  
we assume $\dot{f}\equiv df/ds$ throughout the paper. 
Geodesic motions are 'flat', $\theta=\pi/2$ and the angular momentum is conserved too, giving rise 
to $\dot{\phi}=\textrm{const}\equiv h$, hence $\phi=hs +\phi_0$. In this spacetime each 2-sphere has 
its area equal to $4\pi a^2$. Yet the metric radius of a sphere $t=$const and $x=x_0$, i.~e.~the 
length of the spatial geodesic (in the 3-space $t=$const) from the centre to any point of the sphere is 
$ax_0$. In this sense the variable $x$ is interpreted as a radial coordinate and by a circular worldline 
one means a curve with $x=x_0>0$. The geodesic equation for the variable $x$,
\begin{equation}\label{n70}
\ddot{x}+\dot{t}^2\,\sinh x\cosh x=0,
\end{equation} 
excludes the existence of circular geodesics (with $\theta=\pi/2$ and $\dot{\phi}\neq 0$), hence a body in 
a free fall must approach the centre or recede from it. We shall not consider circular 
worldlines.\\

In the case of the twin A staying at $x=x_0>0$, $\theta=\pi/2$ and $\phi=\phi_0$, the universal integral 
of motion (for $\theta=\pi/2$)
\begin{equation}\label{n71}
g_{\alpha\beta}\dot{x}^{\alpha}\dot{x}^{\beta}= a^2(\dot{t}^2\,\sinh^2x-\dot{x}^2-\dot{\phi}^2)=1
\end{equation} 
implies 
\begin{equation}\label{n72}
t(s)-t_0=\frac{s}{a\sinh x_0}.
\end{equation} 
As in Schwarzschild and CAdS spacetimes gravitation here is attractive and the twin C moves on a radial ($h=0$) 
geodesic as in those cases: at $P_0(t=t_0, x=x_0>0, \phi=\phi_0$) it flies away outwards with initial velocity 
$\dot{x}=u>0$, reaches maximal height $x=x_M$ at $t=t_M$, falls back and returns to the starting place at 
$P_1(t_1=2t_M-t_0, x=x_0)$. For the geodesic C the integral of motion (71) reads, taking into account (69),
\begin{equation}\label{n73}
\dot{x}^2=\frac{1}{a^2}\left(\frac{k^2}{\sinh^2 x}-1\right)
\end{equation} 
\begin{equation}\label{n74}
\textrm{and} \qquad k=(a^2u^2+1)^{1/2}\,\sinh x_0.
\end{equation} 
The highest point of the flight is $\sinh x_M=k$, hence $k>\sinh x_0>0$. Notice that as in anti-de Sitter 
spacetime (\cite{AIS}, \cite{GP} par. 5.2) the spatial infinity is inaccessible for a massive 
particle. In fact, to reach $x=\infty$ it should have infinite energy $k$. Integrating (73) one gets the 
dependence $s(x)$, the expressions are similar to those for CAdS and denoting $\kappa^2\equiv k^2+1$ they read 
\begin{equation}\label{n75}
s(x)=a\left[\arcsin\left(\frac{1}{\kappa}\cosh x\right)-\arcsin\left(\frac{1}{\kappa}\cosh x_0\right)\right]
\end{equation} 
for the outgoing segment ($x$ grows from $x_0$ to $x_M$) and 
\begin{equation}\label{n76}
s(x)=\pi a -a\left[\arcsin\left(\frac{1}{\kappa}\cosh x\right)+\arcsin\left(\frac{1}{\kappa}\cosh x_0\right)\right]
\end{equation}
for $x$ decreasing from $x_M$ to $x_0$ and farther to $x=0$. The length of C from $P_0$ to $P_1$ is 
\begin{equation}\label{n77}
s_C=2s(x_M)=\pi a -2a\,\arcsin\left(\frac{1}{\kappa}\cosh x_0\right)<\pi a.
\end{equation}
For $k\rightarrow\infty$ one gets $s_C\rightarrow \pi a$ for any finite $x_0$. From 
$dt/dx=\dot{t}/\dot{x}$ and relations (69) and (73) one finds the time of flight from 
$P_0$ to $P_1$,
\begin{eqnarray}\label{n78}
\Delta t\equiv t_1-t_0=2(t_M-t_0) & = & 2\ln(k\cosh x_0+\sqrt{k^2-\sinh^2 x_0})-
\nonumber\\
 & - & 2\ln\sinh x_0-\ln(k^2+1).
\end{eqnarray}
We now compare the lengths of worldlines A and C. From (72) the length $s_A$ in the time interval $\Delta t$ is 
\begin{equation}\label{n79}
s_A=a\,\Delta t\,\sinh x_0.
\end{equation}  
A numerical example. For $x_0=5$ and $k=1000$ one gets $x_M=7,60090$, $s_C=2,99304 a$ and $s_A=1,99443 a$, then 
$s_C/s_A=1,50007$; in general there is no doubt that the twin C is older than A at the reunion.

\subsection{Jacobi fields and conjugate points on timelike radial geodesics}
In the $(t,x,\theta,\phi)$ chart the nonvanishing components of the curvature tensor are $R_{0101}=-a^2\,
\sinh^2x$ and $R_{2323}=-a^2\,\sin^2\theta$ and for the Ricci tensor these are $R_{00}=\sinh^2x$, $R_{11}=-1$, 
$R_{22}=+1$, $R_{33}=\sin^2\theta$ and $R=0$. With the aid of the two tensors and the vector $\dot{x}^{\alpha}$ 
tangent to timelike radial ($\theta=\pi/2, \phi=\phi_0$) geodesic curves, which is determined by (69) and (73), 
one finds that these lines contain conjugate points.\\
On the geodesic C a triad of spacelike vector fields satisfying (2) is conveniently chosen as 
\begin{eqnarray}\label{n80}
e_1{}^{\alpha} & = & \left[\frac{\varepsilon (k^2-\sinh^2x)^{1/2}}{a\sinh^2x}, \frac{k}{a\sinh x},
 0,0\right],  
\nonumber\\
e_2{}^{\alpha} & = & \left[0,0,\frac{1}{a},0\right], \qquad 
e_3{}^{\alpha}=\left[0,0,0, \frac{1}{a}\right],
\end{eqnarray}
where $\varepsilon=+1$ on the outgoing segment and $\varepsilon=-1$ on the ingoing one. Employing this 
basis one expands a general Jacobi vector field $Z^{\mu}(s)=\sum_b Z_b(s) e_b{}^{\mu}(s)$, $b=1,2,3$ 
and the geodesic deviation equation for the Jacobi scalars $Z_b(s)$ takes on the following form:
\begin{equation}\label{n81}
\frac{d^2}{ds^2}Z_1+\frac{1}{a^2}\,Z_1=0, \qquad \frac{d^2}{ds^2}Z_2=0 \quad \textrm{and} \quad 
\frac{d^2}{ds^2}Z_3=0.
\end{equation}
These immediately give the generic Jacobi field along the geodesic:
\begin{equation}\label{n82}
Z^{\mu}(s)=(C_{11}\sin\frac{s}{a}+C_{12}\cos\frac{s}{a})\,e_1{}^{\mu}+(C_{21}s+C_{22})\,e_2{}^{\mu}+
(C_{31}s+C_{32})\,e_3{}^{\mu}.
\end{equation}
The special Jacobi scalars vanishing at a given initial point $s=0$ are $Z_1=C_1\,\sin s/a$, $Z_2=C_2\,s$ 
and $Z_3=C_3\,s$. Conjugate points are determined by the special Jacobi field for which $C_2=C_3=0$, then 
the deviation vector $Z^{\mu}=C_{1}e_1{}^{\mu}\sin\frac{s}{a}$ lies in the 2--surface $(t,x)$. 
Assuming that the geodesic infinitely oscillates between the outermost spatial points (as in the CAdS 
spacetime) one finds an infinite sequence of conjugate points $Q_n$ at distances $s_n=n\pi a$, 
$n=1,2,\ldots$, from the initial point. One infers from the spherical symmetry that these points, 
being the cut points, are the only cut points on these curves and there are no other cut points on 
them.  In the case of the geodesic C the nearest point $Q_1$ conjugate to $P_0$ is at the distance 
$s=\pi a$. Since the length (77) of the segment $P_0P_1$ is $s_C<\pi a$, point $Q_1$ is beyond this arc 
and the geodesic C is the longest curve among nearby curves joining $P_0$ and $P_1$, i.~e.~it attains the 
local maximum of length.
 B--R spacetime is not globally hyperbolic and most theorems on maximal curves in the space of all curves 
 joining two given points (see \cite{SG1}) do not apply. By a direct calculation we now show that the 
 ingoing timelike radial geodesics are the maximal curves (their length is equal to the distance function) 
 between any pair of points on the segment from the initial point to a point infinitesimally close to $x=0$ 
 of each geodesic of the class.\\

 To this end we transform from the chart (68) to the Gaussian normal geodesic (GNG) one, i.~e.~comoving 
 coordinates in which the lines of the time coordinate $\tau$ are the radial geodesics. For the reader's 
 convenience we briefly present here the derivation from \cite{SG1} adapted to the B--R spacetime. 
 Usually the GNG chart in a given spacetime is constructed in terms of worldlines of massive particles 
 freely falling down from rest at spatial infinity. In B--R spacetime a particle with finite energy $k$ 
 cannot escape to infinity and according to (73) we assume that a swarm of particles radially falls 
 down from the rest at $x=x_M$, where $\sinh x_M=k$ and $k>0$. Then in the GNG coordinates 
 $(\tau, R, \theta, \phi)$ the velocity field of the radial geodesics is $u^{\alpha}=(1,0,0,0)$ and is 
 the gradient of their common proper time, $u_{\alpha}=(1,0,0,0)=\partial_{\alpha}\tau$. On the other 
 hand the velocity field in the chart (68) has components 
 \begin{displaymath}
 u^{\alpha}=\left[\frac{k}{a\sinh^2x}, -\frac{1}{a}\left(\frac{k^2}{\sinh^2x}-1\right)^{1/2}, 0, 0\right]
 \end{displaymath}  
 and the transformation law for the contravariant components of the field yields
\begin{equation}\label{n83}
\tau=akt+a\int\left(\frac{k^2}{\sinh^2x}-1\right)^{1/2}\,dx.
\end{equation}
In a similar way one gets
\begin{equation}\label{n84}
R=at+ak\int\frac{dx}{\sinh x(k^2-\sinh^2x)^{1/2}}.
\end{equation} 
The inverse transformation is 
 \begin{displaymath}
 \cosh x=\sqrt{k^2+1}\,\sin\left(\frac{kR-\tau}{a}\right).
 \end{displaymath} 
 In the comoving coordinates the B--R metric is 
 \begin{equation}\label{n85}
ds^2=d\tau^2-(k^2-\sinh^2x)dR^2-a^2(d\theta^2+\sin^2\theta\,d\phi^2),
\end{equation} 
here $0\leq \sinh x<k$ and alternatively $-g_{11}=k^2-\sinh^2x=(k^2+1)\cos^2\left(\frac{kR-\tau}{a}\right)$ 
with $-\pi a/2<kR-\tau<\pi a/2$, what implies that the comoving time coordinate is in the interval 
$kR-\pi a/2<\tau<kR+\pi a/2$. Now take any radial geodesic in the domain of the GNG chart, $R=R_0$, 
$\theta=\theta_0$, $\phi=\phi_0$; along it there is $ds=d\tau$. Its length between two points on it, 
$S_1(\tau_1,R_0,\theta_0,\phi_0)$ and $S_2(\tau_2,R_0,\theta_0,\phi_0)$, where $\tau_1$ and $\tau_2$ 
are in the allowed interval, is $\tau_2-\tau_1<\pi a$. Let any 
other timelike curve with the endpoints $S_1$ and $S_2$ be parameterized by $\tau$. Then its length is 
 \begin{eqnarray}\label{n86}
\int^{\tau_2}_{\tau_1}\left[1-(k^2-\sinh^2x)\left(\frac{dR}{d\tau}\right)^2-
a^2\left(\frac{d\theta}{d\tau}\right)^2- a^2\sin^2\theta\left(\frac{d\phi}{d\tau}\right)^2\right]^{1/2}\,
d\tau & &
\nonumber\\
< \tau_2-\tau_1<\pi a. & &
\end{eqnarray}  
Thus in the domain of the comoving chart the radial ingoing geodesics are maximal. From (75) one sees 
that the length of any radial outgoing (or ingoing) timelike geodesic from $x_0$ to $x_M$ is less than 
$\pi a/2$ and tends to this upper limit for $x_M$ and $k$ tending to infinity. This implies that the 
ingoing geodesic from $x_M$ to $x_0$ for any $0<x_0<x_M<\infty$ entirely lies in the chart domain and 
is globally maximal. Since the metric (68) is time symmetric, the same theorem applies to outgoing 
radial geodesics.\\

\section{Conclusions}
 The sample of the three spacetimes considered in this paper as applications of general methods 
 developed in \cite{SG1} do not allow one to formulate a general 
 rule concerning properties of timelike worldlines which may be used in various versions of the twin 
 paradox. On the contrary, even in the maximally symmetric spacetimes, de Sitter and CAdS, one 
 encounters a multitude of posibilities. The physical paradox is reduced to a purely geometrical 
 problem of finding the (possibly unique) longest timelike curve joining two given points. This is a 
 problem in global Lorentzian geometry and it is well known that in globally hyperbolic spacetimes it 
 always has a well defined solution in the form of the maximal timelike geodesic segment whose length 
 is, by definition, equal to the Lorentzian distance function between its endpoints.\\  
  In principle, to find out the maximal geodesic, one must investigate all 
 geodesics between given endpoints. High symmetry of the spacetime is helpful in these investigations 
 to a limited extent. The four spacetimes (including Schwarzschild metric studied in \cite{S}) are 
 spherically symmetric, yet the differences in their global properties are at least as important as 
 their spherical symmetry. Our current study of a general spherically symmetric static spacetime 
 indicates that some common properties of timelike geodesics are accompanied by a diversity of 
 distinct features in various metrics.\\
 
 We choose three physically interesting worldlines: staying at rest and circular and radial motions and 
 solve the twin problem by comparing their lengths. Then we go further and for geodesic worldlines 
 (radial and possibly circular) we determine the geodesic deviation vector fields and conjugate points 
 and in this way we find the locally longest geodesic segments. Finally, in the three spacetimes 
 studied here, de Sitter, CAdS and Bertotti--Robinson, we are able to determine all cut points on the 
 radial and circular geodesics and show that they coincide with the conjugate points.\\
 
 Circular geodesics in covering anti-de Sitter spacetime contain infinite number of conjugate 
 points. While these geodesics lie in the two-surfaces $t-\phi$ ($r=r_0$, $\theta=\pi/2$), 
 the nearby geodesics intersecting them at the conjugate points, lie (besides these points) 
 outside these surfaces.\\
 In CAdS space the radial geodesics (infinitely oscillating between spatial points 
 maximally distant from the centre) contain an infinite sequence of conjugate points equally separated 
 by $\Delta s=\pi a$ and their segments of this length are locally the longest curves between their 
 endpoints. One Jacobi vector field lies in the $t-r$ surface ($\theta=\pi/2$) of the radial geodesics, 
 while the other two fields are directed off it.\\
 Similarly, in Bertotti--Robinson spacetime the oscillating radial timelike geodesics contain 
 infinite number of equally separate conjugate points and these are the only cut points on these 
 curves. Unlike the CAdS case, these points are determined by one deviation vector field, that lying 
 in the two-surface $t-x$.
 
 These few examples clearly show that in dealing with the geodesic deviation vectors and conjugate points 
 one must study case by case.\\
  
  \textbf{Acknowledgements}.\\
We are grateful to Sebastian Szybka for some numerical calculations and for valuable comments. 
 This work was supported by a grant from the John Templeton Foundation.

\end{document}